\documentclass[twocolumn,showpacs,amsmath,amssymb]{revtex4-1}

\usepackage{graphicx}

\begin{document}


\title{Cosmic-Ray Lithium Production at the Nova Eruptions Followed by a Type Ia Supernova}


\author{Norita Kawanaka}
\email{norita@kusastro.kyoto-u.ac.jp}
\affiliation{Department of Astronomy, Graduate School of Science, Kyoto University, Kitashirakawa Oiwake-cho, Sakyo-ku Kyoto, 606-8502, Japan}
\affiliation{Hakubi Center, Yoshida Honmachi, Sakyo-ku, Kyoto 606-8501, Japan}
\author{Shohei Yanagita}
\affiliation{College of Science, Ibaraki University, 2-1-1 Bunkyo, Mito, Ibaraki 310-8512, Japan}

\begin{abstract}
Recent measurements of cosmic-ray light nuclei by AMS-02 have shown that there is an unexpected component of cosmic-ray (CR) lithium whose spectral index is harder than that expected from the secondary production scenario.  We propose the nearby Type Ia supernova following a nova eruption as the origin of lithium nuclei in the CRs.  By fitting the data of CR protons, helium, and lithium fluxes provided by AMS-02 with our theoretical model we show that this scenario is consistent with the observations.  The observational tests that can check our hypothesis are briefly discussed.
\end{abstract}

\date{\today}
\maketitle

\section{Introduction}
It is widely believed that the cosmic-rays (CRs) below $\sim 10^{15.5}~{\rm eV}$ (so called knee) are originated in Galactic supernovae (SNe), which can supply enough energy to maintain the CR energy density in the present day Galaxy and also provide an efficient particle acceleration mechanism by the so-called diffusive shock acceleration \cite{1987PhR...154....1B}.  Actually, recent gamma-ray observations showed that the spectra (in the unit of ${\rm erg}~{\rm cm}^{-2}~{\rm s}^{-1}$) of some supernova remnants (SNRs) below $\sim 200~{\rm MeV}$ are steeply rising and exhibiting a break at $\sim 200~{\rm MeV}$, which indicates that the gamma-ray emissions are produced via hadronic interactions between high energy CR particles accelerated at the SNRs and surrounding gas \cite{2013Sci...339..807A}.

In addition, recent direct observations of CR particles by AMS-02 experiment \cite{2015PhRvL.114q1103A, 2015PhRvL.115u1101A, 2016PhRvL.117w1102A} improve the previous results by PAMELA \cite{2011Sci...332...69A}, CREAM \cite{2010ApJ...715.1400A}, AMS \cite{2000PhLB..490...27A, 2000PhLB..494..193A} , and ATIC-2 \cite{2009BRASP..73..564P}.  Especially, it has clearly shown that the spectra of CR protons and helium nuclei, which are believed to be accelerated at the SNRs primarily, do not have a single power-law form but each of them is dominated by an extra hard component above the energy of $\sim 300~{\rm GV}$.  The origin of this spectral hardening has been controversial: it could be due to the nonlinear acceleration at the source \cite{1997ApJ...487..197E, 2001RPPh...64..429M, 2013ApJ...763...47P}, the energy dependence of the diffusion coefficient \cite{2012Ap&SS.342..131T, 2017arXiv170609812G}, the existence of local sources \cite{2012MNRAS.421.1209T, 2015PhRvL.115r1103K, 2015ApJ...809L..23S, 2017arXiv171002321K}, or the reacceleration of secondary nuclei produced by spallation during propagation \cite{2017MNRAS.471.1662B}.  Moreover, AMS-02 revealed that the spectrum of CR lithium nuclei, which are considered to be produced via nuclear interactions of primary CR nuclei with interstellar matter during propagation \citep{1992ApJ...385L..13S}, has an unexpected hard component which dominates above $\sim 300~{\rm GeV}$ \cite{ting16}.  What is interesting is that the spectral index of the extra CR Li seems quite similar to that of the extra CR protons.  This implies that this extra CR Li component is not produced secondarily (i.e., via the spallation of heavier nuclei), but are accelerated as primary CRs like protons and helium nuclei at the sources \citep{footnote}.

This extra CR lithium may have originated in an environment rich in lithium.  Several sites have been proposed as the source of lithium in the Universe: the Big bang nucleosynthesis (see \cite{2016RvMP...88a5004C} for review), stellar flares in evolved low mass stars \cite{1995ApJ...453..810D}, supernova explosions \cite{1978Ap&SS..58..273D}, and novae \cite{1978ApJ...222..600S}.  Direct observational evidence of lithium production in these sites have not been found for a long time, but very recently, Tajitsu et al.\cite{2015Natur.518..381T} reported the detection of $^7 {\rm Be}$ in the post-outburst spectra of the classical nova V339 Del (Nova Delphini 2013), which shows that $^7{\rm Be}$ nuclei are created during the nova explosion via $\alpha$-capture reaction, $^3{\rm He}(\alpha,\gamma)^7{\rm Be}$.  Since $^7{\rm Be}$ decays to $^7{\rm Li}$ with half-life of $53.22~{\rm days}$, this observation strongly supports that novae supply significant amount of $^7{\rm Li}$ to the interstellar medium.  They evaluated the mass fraction of $^7{\rm Be}$ in the ejecta as $X(^7{\rm Be})\sim 10^{-4.3\pm 0.3}$.  These discoveries show that novae are important production sites of $^7$Li nuclei (see also \cite{2015ApJ...808L..14I, 2016MNRAS.463L.117M, 2016ApJ...818..191T}).  

Nova eruption occurs when the hydrogen gas accreted onto a white dwarf from its companion star accumulates and undergoes thermonuclear runaways.  This happens when the mass accretion rate is as low as $\dot{M}\lesssim 10^{-8}-10^{-7}~M_{\odot}~{\rm yr}^{-1}$.  In a single nova eruption the ejecta mass would be up to $M_{\rm ej, nova}\sim 10^{-4}M_{\odot}$ \cite{1980AJ.....85..283S}, and it would be mixed with the interstellar medium to supply significant amount of lithium ($\gtrsim 10^{-8}M_{\odot}$ per nova eruption) into the interstellar space.   If the accretion rate onto a white dwarf from its non-degenerate companion star is above the mass range which is so called the stability strip ($\dot{M}\sim 5\times 10^{-7}~M_{\odot}~{\rm year}^{-1}$ for a $1.4~M_{\odot}$ white dwarf; \cite{1982ApJ...253..798N}), a white dwarf gains mass toward the Chandrasekhar limit thanks to stable hydrogen burning on its surface, which may result in the thermonuclear disruption.  This is the so called single degenerate scenario of Type Ia supernovae \cite{1973ApJ...186.1007W, 2012NewAR..56..122W} (see also \cite{2014ARA&A..52..107M}).  It is proposed that recurrent novae, which have undergone a number of eruptions, are one of the possible channels for Type Ia supernova explosions (e.g. \cite{2000ApJ...536L..93H, 2001ApJ...558..323H}).  If this is true, the circumstellar matter polluted by a number of preceding nova ejecta rich in lithium, should surround the progenitor of the Type Ia supernova and could supply plenty of lithium as seed nuclei of CRs.  Actually, a recently discovered supernova PTF11kx showed saturated Ca II H and K absorption lines and weak Na I D lines in its initial spectrum, and a strong H$\alpha$ line emission with a P-Cygni profile in its subsequent emission \cite{2012Sci...337..942D}.  These features imply that this supernova is interacting with multiple expanding shells, and it is suggested that PTF11kx was a bona fide Type Ia supernova with a nova progenitor.

We propose that the CR lithium nuclei would be efficiently accelerated when the nova ejecta is swept up by a blast wave from a subsequent Type Ia supernova explosion (see Fig.1), and that the extra hard components of CR protons, helium, and lithium are produced at a nearby Type Ia supernova occurring in such an environment.  In the following sections, we investigate quantitatively the possibility that a nearby Type Ia supernova occurring after the nova eruption is the origin of observed CR lithium excess, and make some predictions for the future CR experiments.

\begin{figure}[tbp]
\begin{center}
\includegraphics[width=3.4in]{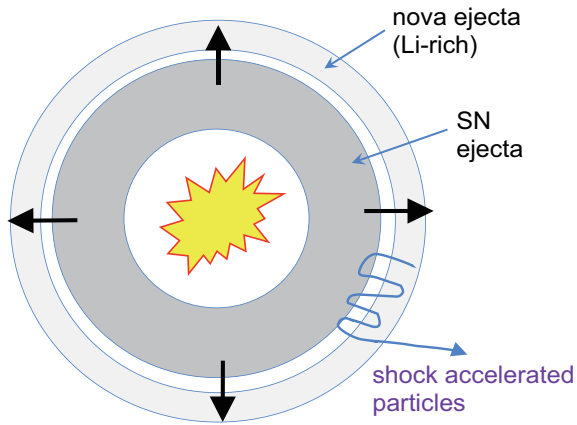} 
\caption{Schematic picture of a Type Ia supernova occurring after a nova eruption from a white dwarf.  Since the nova ejecta is Li-rich, the subsequent supernova blast wave can accelerate a significant amount of lithium nuclei.}
\label{fig1}
\end{center}
\end{figure}

\section{Model}
We assume that the flux of proton, helium, and lithium observed by AMS-02 is a superposition of two components; one is from a hypothetical nearby Type Ia supernova and the other is the background flux due to myriad sources in the Galaxy (e.g. \cite{2003ApJ...582..330H}) except the hypothetical supernova.  We deduce some properties of the hypothetical supernova in the framework of a simple propagation model from the extra CR components which is estimated by the observed flux from which the background component is subtracted.  The background fluxes of CR protons and helium as functions of rigidity $R$ can be well fitted by a single power-law with indices of $\sim -2.849$ \cite{2015PhRvL.114q1103A}, $\sim -2.780$ and \cite{2015PhRvL.115u1101A}, respectively.  We can also fit the background flux of CR lithium with a single power-law with index $\sim -3.1$ (see Fig. 2).  Above $\sim 300~{\rm GeV}$, each of CR fluxes is dominated by an extra component with a harder index.  Let us account for these extra components with the contribution from a single supernova remnant.  The propagation of CR nuclei in the interstellar medium can be described by the diffusion equation,
\begin{eqnarray}
\frac{\partial }{\partial t}f_i(r, \epsilon, t)=D(R)\nabla^2 f_i +Q_i(r,R,t),
\end{eqnarray}
where  $f_i(r,R,t)$ is the distribution function of CR $i$-particles at the distance $r$ from a source and the time of $t$, with rigidity of $R$, $D(R)$ is the diffusion coefficient in the interstellar medium, and $Q_i(r,R,t)$ is the CR injection rate from a source.  Here we assume that the propagation is described as a pure diffusion process and that the energy losses and nuclear reactions are neglected.  When we assume the instantaneous CR injection from a point-like source (i.e., $Q_i \propto \delta(t) \delta (r)$), the solution of this equation can be described as
\begin{eqnarray}
f_i(r,R,t)=\frac{Q_{i,0}(R)}{(4\pi Dt)^{3/2}}\exp \left( -\frac{r^2}{4Dt} \right), \label{greenfunction}
\end{eqnarray}
where $Q_{i,0}(R)$ is the injection spectrum of CR particles $i$.  It is often assumed that the diffusion coefficient has a power-law like dependence on energy: $D(R)=D_0 (R/1~{\rm GV})^{\delta}$, where $D_0$ and $\delta$ are constants.  Hereafter we assume the injection spectrum, $Q_{i,0}$, as a single power-law function of the energy $\epsilon=\sqrt{(Z_i e R)^2+(N_i m_pc^2)^2}$ with index of $\alpha$ (i.e. $Q_{i,0}(R)=q_{i,0}\epsilon^{-\alpha}$ where $q_{i,0}$ is the normalization factor), where $Z$ and $N$ are the atomic number and mass number of a particle $i$, respectively.  In this case Eq. (\ref{greenfunction}) can be approximated by a power-law with index of $\alpha +(3/2)\delta$ in the energy range where the diffusion length $(4Dt)^{1/2}$ is much larger than the distance $r$.

The distribution function shown in Eq. (\ref{greenfunction}) is a convex function of rigidity which attains its maximum value at
\begin{eqnarray}
R_{\rm p}= \left [ \frac{\delta}{\alpha +\frac{3}{2}\delta} ~\frac{r^2}{r_0^2} \right]^{1/\delta},
\end{eqnarray}
where $r_0 = (4D_0 t)^{1/2}$ is the diffusion length of a particle with rigidity of $1~{\rm GV}$.  Assuming that the rigidity index of the diffusion coefficient is $\delta=1/3$, which has been indicated by the recent AMS-02 result \cite{2016PhRvL.117w1102A}, this peak rigidity can be described as
\begin{eqnarray}
R_{\rm p}&\simeq &195~{\rm GV} \left( \frac{r}{250~{\rm pc}} \right)^6 \left( \frac{t}{10^4~{\rm yr}} \right)^{-3} \nonumber \\
&&\times \left( \frac{D_0}{10^{28}~{\rm cm}^2~{\rm s}^{-1}} \right)^{-3}, \label{peak}
\end{eqnarray}
where we fix the spectral index at the source as $\alpha=2.2$ so that the spectral index above the peak is nearly equal to that of the observed extra flux component, $\gamma+\Delta \gamma \simeq -2.7$.  In order to fit the hard extra components by Eq. (\ref{greenfunction}), this peak rigidity should be much lower than $\sim 300~{\rm GV}$, above which the extra component can be approximated by a simple power-law without a break.

\section{Results and Discussion}

By fitting the data of CR protons, He and Li by AMS-02 with our model spectra, we can evaluate the distance and age of the hypothetical source and the normalization factors $q_{i,0}$ ($i=p,~{\rm He}~{\rm and}~{\rm Li}$) using Levenberg-Maruardt method.  Fig. 2 depicts the spectra of CR protons, helium and lithium fluxes as a function of rigidity by AMS-02, fitted by our model.  The hard components appearing above $\sim 300~{\rm GV}$ are fitted by our Type Ia supernova model using parameters of $D_0=1\times 10^{28}~{\rm cm}^2~{\rm s}^{-1}$, $r=150~{\rm pc}$, $t=6\times10^3~{\rm year}$, and the mass of each kind of CR particles is $M_{{\rm CR},p}=2.0\times 10^{-6}M_{\odot}$, $M_{\rm CR,He}=1.3\times 10^{-6}M_{\odot}$, and $M_{\rm CR,Li}=1.0\times 10^{-8}M_{\odot}$.   Here the mass of each CR component is obtained by
\begin{eqnarray}
M_{{\rm CR},i}=N_i m_p c^2 \times \int_{N_i m_pc^2}^{\epsilon_{\rm max}}d\epsilon~ Q_{i,0}(R),
\end{eqnarray}
where $\epsilon_{\rm max}$ is the upper limit of this energy integration.  Since we assume that the injection spectrum $Q_{i,0}(R)$ has a spectral index steeper than $2$, the choice of $\epsilon_{\rm max}$ does not affect the results so much.  Note that the different choice of the rigidity index of the diffusion coefficient $\delta$ would impact the fitting parameters such as the age and distance of the source by a factor of a few at most.

One can easily see from the equation (\ref{greenfunction}), the functional form of the extra component would be the same as long as the ratios $Q_{i,0}(R)/(D_0t)^{3/2}$ and $r^2/(D_0t)$ are unchanged.  This means that the different choice of the diffusion coefficient $D_0$ gives us the different parameter set that can fit the observed CR data.  Even if $D_0$ is fixed, one cannot determine uniquely the distance and age of the CR source, and the total CR masses.  However, we should note that, with the condition that the peak energy of the extra component (\ref{peak}) is much lower than $\sim 300~{\rm GV}$ and that the total CR energy $\sum_i E_{{\rm CR},i}$ is smaller than the typical energy injected into CRs per supernova $\sim 10^{50}~{\rm ergs}$, we can give the constraint on the distance of the source as
\begin{eqnarray}
r \lesssim 350~{\rm pc},
\end{eqnarray}
which is independent of the diffusion coefficient.  In other words, the source of the extra CR components should be a local source within our model.

\begin{figure}[t]
\begin{center}
\includegraphics{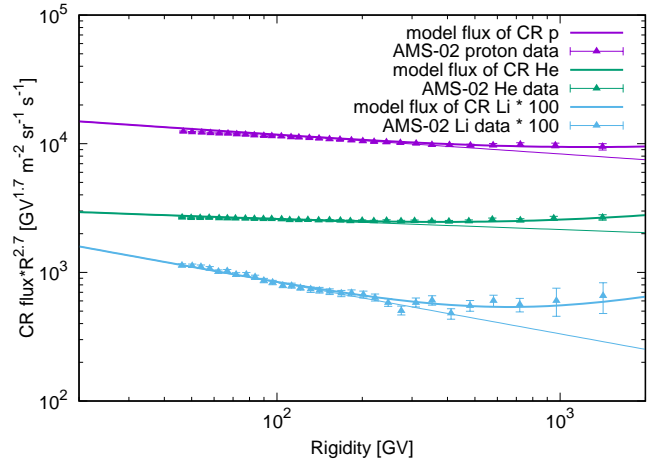}
\caption{Comparison of our model spectra of CR protons, helium and lithium nuclei with AMS-02 data.  The data points are given by AMS-02 (\cite{2015PhRvL.114q1103A} for protons, \cite{2015PhRvL.115u1101A} for helium, and \cite{ting16} for lithium).
}
\label{fig2}
\end{center}
\end{figure}

Let us discuss the relevance of the parameters adopted in the fitting presented in the previous section.  In the example presented in the previous section it is required that there is only a single source that emit CR protons, helium and lithium (i.e., a SN Ia occurring after nova eruptions) in the vicinity of the solar system (within a few hundred parsecs) in the last $\sim 10^4~{\rm years}$.  The rate of SNe Ia in our Galaxy is estimated as $\simeq 1.4$ per century for the total Galactic supernova rate of $\simeq 4.6$ per century \cite{2013ApJ...778..164A}.  In addition, it is observationally implied that the fraction of SNe Ia that show strong interaction with their circumstellar medium like PTF11kx is $\sim 20\%$ of all SNe Ia \cite{2013MNRAS.436..222M}.  From these facts, we can roughly estimate the rate of SNe Ia that have strong interaction with circumstellar medium in the vicinity of the solar system ($\lesssim 1~{\rm kpc}$) as no more than $\lesssim 0.03$ per century, and then our assumption is marginally reasonable.

The relevance of the amount of CRs emitted from a source should also be checked.  In our model, the total mass of CR particles coming from our hypothetical source is $\sum_i{M_{{\rm CR},i}}\simeq 3.3\times 10^{-6}~M_{\odot}$.  Comparing this value to the typical mass of ejecta erupted from a nova eruption, $\sim 10^{-4}~M_{\odot}$ \cite{seaquistbode08}, we obtain the required acceleration efficiency in our scenario would be up to $\sim 3\times 10^{-2}$.  According to \cite{2016AdSpR..57..519E}, who investigated the fraction of particles in the ambient plasma that are accelerated by the SNR shocks as a function of temperature, the acceleration efficiency would be greater than a few times $10^{-2}$ when the temperature is higher than $\gtrsim 10^4~{\rm K}$.  According to the radio observations the temperature of nova ejecta is about $\gtrsim 10^4~{\rm K}$ \cite{1979AJ.....84.1619H, 2014MNRAS.442..713M}, which would justify the large acceleration efficiency required in our model.  Moreover, if there were multiple nova eruptions prior to the supernova explosion, which is just what was implied by the observation of PTF 11kx \cite{2012Sci...337..942D}, the required acceleration efficiency would be relaxed.

The abundance ratios of CR particles in the extra components are also important.  Our fitting shows that the mass ratio of helium to protons ($M_{\rm CR, He}/M_{{\rm CR},p}$), and the ratio of lithium to protons in CR particles ($M_{{\rm CR, Li}}/M_{{\rm CR},p}$) in the extra hard component are $\simeq 0.65$ and $\simeq 0.005$, respectively.  The former ratio is consistent with the theoretical calculations of nucleosynthesis in the nova ejecta (e.g. \cite{1998ApJ...494..680J, 2016PASJ...68...39L}).  The latter ratio is an order of magnitude larger than the mass ratio of lithium in the nova ejecta inferred from the observation, $\sim 10^{-4}$ \cite{2015Natur.518..381T, 2016ApJ...818..191T}.  However, this gap would be filled up by taking into account the dependence of acceleration efficiency on the first ionization potential (FIP).  Meyer shows how the overabundance of elements in Galactic CRs with respect to local interstellar medium could be related to their FIPs \cite{1985ApJS...57..173M}.  According to his results, the abundance of sodium, whose FIP is $\sim 5~{\rm eV}$, in the Galactic CRs is similar to that in the interstellar medium, while the abundance of hydrogen, whose FIP is $\sim 13~{\rm eV}$, in the Galactic CRs is a factor of $\sim 0.03$ smaller than that in the interstellar medium.  This implies that the elements with smaller FIPs are more likely to be accelerated at the shock than those with larger FIPs.  Since the FIP of lithium is about $\sim 5~{\rm eV}$, the mass fraction of lithium in the Galactic CRs would be enhanced by a factor of a few tens from that in the nova ejecta.  If this is the case, the abundance ratio in the extra component obtained by the fitting is consistent with our model.  The volatility of elements is the other relevant atomic property that may control the enhancement of some elements in CRs \cite{1997ApJ...487..182M}.  According to \cite{2003ApJ...591.1220L}, the 50\% condensation temperature of Li is as high as $1142~{\rm K}$, which is comparable with that of P ($1229~{\rm K}$).  This implies that Li is as refractory as P, and that its acceleration efficiency would be enhanced by a factor of $O(10)$ with respect to volatile elements, such as protons and helium.  As for the abundance and spectral indice of CRs below the break, our scenario does not give any implication because in that energy range the contribution from our hypothetical supernova is subdominant.  To account for the CR spectra and composition below $\sim 300~{\rm GV}$, we can consider the CR acceleration at supernovae in superbubbles \citep{2011ApJ...729L..13O, 2016PhRvD..93h3001O}.

As CR protons and helium, CR lithium also show a spectral break at $\sim 300~{\rm GV}$ as we mentioned above \citep{yan+17}.  These three elements show their spectral breaks at remarkably similar rigidity values regardless of the large difference in the intensities.  However, this similarity may only be accidental one when we consider that lithium in the background Galactic CR (i.e. in the rigidity range below the break) is purely secondary CRs in contrast to the primary origin of protons and helium.

Finally we would like to emphasize that our scenario presented in this letter predicts some interesting specific properties of GCRs which could be checked by direct observations in space or with balloons such as AMS-02, CALET \cite{2015ICRC...34..581T}, DAMPE \cite{2017EPJWC.13602010D}, ISS-CREAM \cite{isscream}, and others including future experiments.  First, the fluxes of CR beryllium and boron nuclei would not show any excesses over the component expected from spallation of heavier elements in the energy range above $\sim 300~{\rm GV}$ because these two elements would not be synthesized in nova explosions.   Second, CREAM reported a hint of a spectra hardening in CR carbon flux as well as in the other major heavy elements up to iron above $\sim 200~{\rm GeV}/{\rm n}$ \cite{2012APh....39...76S}.  Recent precise measurements of CR carbon, nitrogen, and oxygen fluxes by AMS-02 also revealed a surprising spectral hardening above $\sim 300~{\rm GV}$, almost the same rigidity where the spectral hardening was seen for proton, helium, and lithium \cite{yan+17}.  The spectral hardening of the heavy elements including carbon may be naturally expected in our scenario, because significant amount of these elements could have been synthesized in the nova explosions and could be accelerated in the subsequent Type Ia supernova together with lithium, proton and helium of the same origin.  On the other hand, a spectral hardening should not be seen in the boron flux, because boron cannot be synthesized in the nova explosions as stated above.  Consequently the boron to carbon ratio would decrease much faster with energy than expected from a naive model for the energy dependence of the escape time of GCRs from the Galaxy, although in the paper \cite{2016PhRvL.117w1102A} they claim that the new AMS-02 data of the ratio up to $\sim 2.6~{\rm TV}$ reveals that its energy dependence is described by a single power-law with the index expected from the Kolmogorov model of the interstellar turbulence.  The departure of the value expected by our model from this simple power-law would be seen clearly in the much higher energy range than covered by AMS-02 observations.  Third, the energy dependence of the isotope ratio $^7{\rm Li}/^6{\rm Li}$ would be the most crucial test for our scenario.  In the lower energy range below $\sim 300~{\rm GV}$, both isotopes are mainly the spallation products of the background heavier CR elements such as carbon and oxygen.  Accordingly the isotopic ratio is determined essentially by the ratio of the spallation cross section value of each isotope modified by the solar modulation effect.  On the other hand in the higher energy range above $\sim 300~{\rm GeV}$, this isotopic ratio would dramatically increase with energy because this extra flux of lithium from the hypothetical supernova should consist of almost pure $^7{\rm Li}$, because only $^7{\rm Li}$ isotope should have been synthesized in the nova explosions which were followed by the supernova explosion as stated in the introduction.  Although the separation of these isotopes at these energies would be impossible in the current experiments such as AMS-02, it is one of the promising probe to distinguish models by future experiments.  Finally, we point out a possibility of an energy dependent, strong dipole anisotropy in the CR arrival direction, because the source of the extra component is estimated to be rather close to the solar system and to be relatively young.  Observations of the anisotropy by various instruments will allow another test of our scenario of nearby Type Ia supernova model and also may give important constraints on the value of the local diffusion coefficient, because the degree of expected anisotropy strongly depends on the absolute value of the diffusion coefficient.  The details of our predictions stated above will be discussed more quantitatively in the forthcoming paper.
\begin{acknowledgements}
We acknowledge Anonymous Referees for their quite helpful, critical comments.  We would like to thank Satoru Katsuda and Shiu-Hang Lee for their helpful comments.  This work was supported by Hakubi project at Kyoto University.
\end{acknowledgements}


\end{document}